\documentclass[amsmath,amssymb,aps,prl,twocolumn]{revtex4-2}
\usepackage{txfonts}
\usepackage{bm}
\usepackage{graphicx}

\newcommand{\be}{\begin{eqnarray}}
\newcommand{\ee}{\end{eqnarray}}
\newcommand{\bmat}{\left(\begin{array}}
\newcommand{\emat}{\end{array}\right)}
\newcommand{\no}{\nonumber}
\newcommand{\diff}{\mathrm d}
\newcommand{\e}{\mathrm e}

\begin{document}

\title{Nonadiabatic Control of Geometric Pumping}
\author{Kazutaka Takahashi}
\affiliation{Institute of Innovative Research, Tokyo Institute of Technology, Kanagawa 226--8503, Japan}

\author{Keisuke Fujii}
\affiliation{Department of Physics, Tokyo Institute of Technology, Tokyo 152--8551, Japan}
\affiliation{iTHEMS Program, RIKEN, Saitama 351--0198, Japan}
\author{Yuki Hino}
\affiliation{Yukawa Institute for Theoretical Physics, Kyoto University, Kyoto 606--8502, Japan}
\author{Hisao Hayakawa}
\affiliation{Yukawa Institute for Theoretical Physics, Kyoto University, Kyoto 606--8502, Japan}

\date{\today}

\begin{abstract}
We study nonadiabatic effects of geometric pumping.
With arbitrary choices of periodic control parameters,
we go beyond the adiabatic approximation to obtain the exact pumping current.
We find that a geometrical interpretation for the nontrivial part 
of the current is possible even in the nonadiabatic regime.
The exact result allows us to find a smooth connection between  
the adiabatic Berry phase theory at low frequencies
and the Floquet theory at high frequencies.
We also study how to control the geometric current.
Using the method of shortcuts to adiabaticity
with the aid of an assisting field, 
we illustrate that it enhances the current.
\end{abstract}

\maketitle

{\it Introduction}. 
In 1983, Thouless discovered a phenomenon called geometric pumping.
In electron systems, a slow periodic variation of control parameters 
gives a nontrivial current without bias~\cite{Thouless,NT}.
The mechanism is described by the geometric Berry phase~\cite{Berry84},
which shows that it is a topological phenomenon.
While the original study was applied to a one-dimensional system
with a lattice potential,  
we can find various processes driven by Berry phase 
in mesoscopic quantum dot systems~\cite{Brouwer},
and in stochastic systems described 
by the classical master equation~\cite{Parrondo,ULB,Astumian03,Astumian07,SN07-1,SN07-2,RHJ,CKS12-1,CKS12-2} 
or the quantum master equation~\cite{RB,BV,CAP,YSSH12,YSSH13,NTKT,NT17}.
The experimental verification can be seen 
in many works~\cite{PLUED,SMCG,Betal,Ketal,XCN,Netal,LSZAB,Metal}.
The pumped system is also interesting from 
a viewpoint of stochastic thermodynamics.
In small systems with appreciable fluctuations, 
by using the method of full counting statistics~\cite{LL,LLL,BN},
we can examine the fluctuation theorem~\cite{ECM,GC,SU,SH}.

Although the phenomenon is a purely dynamical one, 
the theoretical description relies on the static picture.
The use of the adiabatic approximation is crucial 
not only for theoretical analysis 
but also for establishing the geometrical picture.
Since the adiabatic approximation is justified only 
at the case when the parameter change is sufficiently slow, 
it is important to ask 
how much the adiabatic description makes sense 
for nonideal fast manipulations.
It is known that the geometric phase for nonadiabatic systems 
is still useful~\cite{AA,SB,BMKNZ}, but 
we have not fully understood the corresponding phenomenon  
for the geometric pumping.
A breakdown of the fluctuation theorem in the adiabatic regime 
was reported in~\cite{RHL, WH17, HH} and 
it is an interesting problem 
to study how the nonadiabatic effect changes the result.
While nonadiabatic effects in the geometric pumping have been studied
in many works~\cite{Ohkubo08-1,Ohkubo08-2,CGK,Ohkubo13,Uchiyama,WH14,PRCS},
we need a nonperturbative analytical method 
to obtain a clear picture of the nonadiabatic pumping.
Establishing the nonadiabatic description is important
not only for finding the fundamental properties
but also for realizing efficient control of systems in applications.

In this letter, we treat the stochastic master equation 
to study the nonadiabatic effect.
We propose a method incorporating the effect 
to the solution of the equation.
We find that a geometrical interpretation is still possible for
the pumping current under modulation with arbitrary speed, 
which allows us to discuss controlling 
the nontrivial contributions to the current.

{\it Master equation}.
The system we treat in this letter is coupled to 
several reservoirs to provide particle transfer.
The process is stochastic and 
the time evolution of the system is described by 
the master equation 
\be
 \frac{\diff}{\diff t}|p(t)\rangle = W(t)|p(t)\rangle. \label{master}
\ee
$|p(t)\rangle$ is represented as $|p(t)\rangle=(p_1(t),p_2(t),\dots)^{\rm T}$ 
where the $i$th component represents the probability of the $i$th microscopic state
of the system being occupied.
$W(t)$ is a transition-rate matrix with 
each component $W_{ij}(t)$ representing the transition rate from 
state $j$ to state $i$ at $t$.
The system is coupled to reservoirs and 
$W(t)$ is decomposed as $W(t)=\sum_\nu W^{(\nu)}(t)$ 
where $\nu$ labels the reservoirs.
$W_{ij}^{(\nu)}(t)$ is defined in a similar way.
The off diagonal components of $W^{(\nu)}(t)$ are nonnegative
and the diagonal components must satisfy the condition
$\sum_iW_{ij}^{(\nu)}(t)=0$. 
To find a nontrivial contribution to the current, 
we modulate the system periodically without the average bias
between the left ($\nu={\rm L}$) and right ($\nu={\rm R}$) couplings.  

Assuming that the transition-rate matrix is diagonalizable, 
we represent the solution of the master equation 
by an orthonormal set of the instantaneous
left and right eigenstates of $W(t)$, 
denoted as $\{\langle\phi_n(t)|, |\phi_n(t)\rangle\}$
with the eigenvalues $\{\epsilon_n(t)\}$ where $n$ is the index specifying
the corresponding eigenvalue. 
Since the transition-rate matrix is non-Hermitian,
the left eigenstate is not equal to the conjugate of the right eigenstate.
See Supplemental Material (SM) for details.
We write  
\be
 &&|p(t)\rangle = \sum_n C_n(t)
 \e^{\int_0^t \diff t'\,\epsilon_n(t')}|\tilde{\phi}_n(t)\rangle, \label{p0} \\
 && |\tilde{\phi}_n(t)\rangle=\e^{-\int_0^t \diff t'\,\langle\phi_n(t')|\dot{\phi}_n(t')\rangle}|\phi_n(t)\rangle,
\ee
where the dot denotes the time derivative.
$|\tilde{\phi}_n(t)\rangle$ represents the eigenstate with a geometric ``phase''
which is an analog of the Berry phase, or the Aharonov--Anandan phase, 
in quantum mechanics~\cite{Berry84,AA,SB,BMKNZ}.
This state vector has the property of the gauge invariance,
that is the invariance under the transformation
$(\langle \phi_n(t)|,|\phi_n(t)\rangle)\to (\langle\phi_n(t)|R_n^{-1}(t),R_n(t)|\phi_n(t)\rangle)$ where $R_n(t)\in \mathbb{R}$ with $R_n(0)=1$.
To find the geometric current, we use the adiabatic approximation, namely, 
the time dependence of the coefficients $C_n(t)$ is neglected. 
The physical meaning of this approximation is that 
the system follows an instantaneous eigenstate 
of the system when the time variation of $W(t)$ is small.
To examine effects of fast driving, we need to treat mixing between 
different eigenstates.

The master equation has, at least, one stationary state with zero eigenvalue.
For simplicity, we assume that 
this stationary state, denoted with the label $n=1$, is unique.
Then, $C_1(t)=1$ and the other states with $n\ne 1$ 
have negative eigenvalues $\epsilon_n(t)<0$.
The equation for $C_n(t)$ with $n\ne 1$ is given by 
\be
 \frac{\diff C_n(t)}{\diff t}\e^{\int_0^t\diff t'\,\epsilon_n(t') }
 +\sum_{m(\ne n)} C_m(t)\e^{\int_0^t\diff t'\,\epsilon_m(t') }
 \langle \tilde{\phi}_n(t)|\dot{\tilde{\phi}}_m(t)\rangle = 0.
 \label{eqc}
\ee
When we consider a slow modulation, we expect that 
the time evolution does not make transitions to different eigenstates.
This means that the overlap in the second term
on the left hand side of Eq. (\ref{eqc}), 
$\langle \tilde{\phi}_n(t)|\dot{\tilde{\phi}}_m(t)\rangle
=\langle \tilde{\phi}_n(t)|\dot{W}(t)|\tilde{\phi}_m(t)\rangle/(\epsilon_m(t)-\epsilon_n(t))$ with $m\ne n$, is negligible.
In addition, in systems described by the master equation,
we have an exponentially decaying factor
$\e^{\int_0^t\diff t'\,\epsilon_m(t') }$ for $m\ne 1$,
which further justifies the approximation.
The factor is absent for $m=1$ with $\epsilon_1(t)=0$ 
and it is reasonable to keep this term.
Then, neglecting the contributions with $m\ne 1$, 
we obtain a nonadiabatic approximate solution
\be
 |p(t)\rangle \simeq |\tilde{\phi}_1(t)\rangle+\sum_{n\ne 1}
 \left(\delta_n(t)+C_n\e^{\int_{0}^{t} \diff t'\,\epsilon_n(t')}\right)
 |\tilde{\phi}_n(t)\rangle, \label{p}
\ee
where $C_n$ is a constant determined from the initial condition, and 
\be
 \delta_n(t) = -\int_0^t \diff t'\,
 \langle\tilde{\phi}_n(t')|\dot{\tilde{\phi}}_1(t')\rangle
 \e^{\int_{t'}^{t} \diff t''\,\epsilon_n(t'')}. 
\ee
See SM for details of the derivation.
The adiabatic approximation for $|p(t)\rangle$ 
is obtained by setting $\delta_n(t)=0$.
$\delta_n(t)$ depends on
the whole history of the time evolution and represents nonadiabatic effects.
This function is not periodic in $t$ even when 
$W(t)$ is periodic.
However, it rapidly falls into a periodic behavior 
at large $t$.
$\delta_n(t)$ falls into the same trajectory  
after transient evolutions at first several periods 
(See SM).
A similar function appears in quantum systems
to treat a nonreciprocal effect for Landau-Zener tunneling~\cite{KNM},
where the function was evaluated by using a contour integral 
in a complex plane.

{\it Pumping current}.
Using the solution of the master equation (\ref{master}), Eq.~(\ref{p}), 
we can evaluate the current through the system.
Formally, it can be defined by introducing a counting field~\cite{SN07-1}.
To make the discussion concrete, 
we treat the two-state case where the number of the components 
of $|p(t)\rangle$ is two and Eq.~(\ref{p}) becomes the exact solution.
When we set that the first (second) component of $|p(t)\rangle$ represents 
the probability that the system is empty (filled), 
the average current through the system from 
the left to right reservoirs is given by 
$J=\lim_{T\to\infty}\frac{1}{T}\int_0^T \diff t\, 
\left(W_{12}^{\rm (R)}(t)p_2(t)-W_{21}^{\rm (R)}(t)p_1(t)\right)$ (See SM).
In this expression, 
the long-time averaged current is independent of the initial condition and 
of the last term in the brackets of Eq.~(\ref{p}).
This implies that we can calculate the exact current 
by using the approximated state in Eq.~(\ref{p})
even if we go beyond the two-state case.
The neglected term in Eq.~(\ref{eqc}) incorporates
an exponentially-decaying factor and does not contribute to the current
after the second modulation cycle.

In the adiabatic approximation for the current, 
$J$ is given by the sum of the dynamical part $J_{\rm d}$ and
the geometric part $J_{\rm g}$.
The former is given by the dynamical ``phase'' term and 
the latter by the geometric term~\cite{SN07-1}.
In the present treatment, 
the dynamical part is the same and 
the geometric part is separated into the adiabatic part and 
the nonadiabatic part $J_{\rm g}=J_{\rm ad}+J_{\rm nad}$.
The explicit form of each part is respectively given by 
\be
 &&J_{\rm d}
 =\frac{1}{T_0}\int_0^{T_0} \diff t\,
 \frac{k_{\rm in}^{\rm (L)}(t)k_{\rm out}^{\rm (R)}(t)-k_{\rm out}^{\rm (L)}(t)k_{\rm in}^{\rm (R)}(t)}{k_{\rm in}(t)+k_{\rm out}(t)}, \\
 && J_{\rm ad}
 = \frac{1}{T_0}\int_0^{T_0} \diff t\, p^{\rm (R)}(t)
 \frac{\diff}{\diff t}p_{\rm out}(t), \label{Jad}\\
 && J_{\rm nad}
 = \lim_{T\to\infty}\frac{1}{T_0}\int_{T}^{T+T_0} \diff t\,p^{\rm (R)}(t)
 \frac{\diff}{\diff t}\delta_2(t), \label{Jnad}
\ee
where we put $W_{12}(t)=k_{\rm out}(t)=k_{\rm out}^{\rm (L)}(t)+k_{\rm out}^{\rm (R)}(t)$,
$W_{21}(t)=k_{\rm in}(t)=k_{\rm in}^{\rm (L)}(t)+k_{\rm in}^{\rm (R)}(t)$, and 
$p^{\rm (R)}(t)=(k_{\rm in}^{\rm (R)}(t)+k_{\rm out}^{\rm (R)}(t))/(k_{\rm in}(t)+k_{\rm out}(t))$, 
$p_{\rm out}(t)=k_{\rm out}(t)/(k_{\rm in}(t)+k_{\rm out}(t))$.
Here, $k_{\rm in}(t)$ represents the incoming rate and 
$k_{\rm out}(t)$ the outgoing rate, and the superscript denotes
the coupling to the left or right reservoir.
We consider the case where each parameter is represented 
as a function of $\omega t$ with the period $T_0=2\pi/\omega$.
The dynamical part is independent of $\omega$
and is negligible for no-biased pumping.
$J_{\rm ad}$ is represented by using the geometric term 
and is proportional to $\omega$.
Therefore, within the adiabatic approximation, 
the current is enhanced by increasing $\omega$, 
though the expression is only valid in the limit $\omega\to 0$.
This behavior is interfered by the presence of $J_{\rm nad}$.
We stress that the above form of the current is exact.
By knowing the explicit form of the nonadiabatic part,
we can optimize the current as we discuss below.
It is a straightforward task to find a similar expression of the current 
in general multilevel systems.

\begin{center}
\begin{figure}[t]
\begin{center}
\includegraphics[width=0.8\columnwidth]{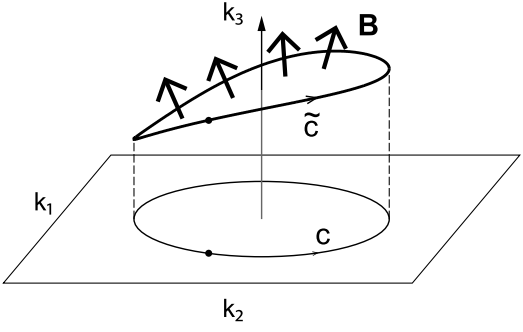}
\end{center}
\caption{
Trajectories in the parameter space.
When we consider a periodic trajectory C in the $(k_1,k_2)$ plane, 
$k_3$ is changed accordingly and we have a closed contour $\tilde{\rm C}$. 
The current is determined by the magnetic field penetrating
a surface $\tilde{S}$ specified by $\tilde{\rm C}=\partial\tilde{S}$.
}
\label{fig1}
\end{figure}
\end{center}

{\it Geometrical picture}.
The nonadiabatic part, Eq.~(\ref{Jnad}), has a similar form 
to the adiabatic part, Eq.~(\ref{Jad}),
which leads to a geometrical interpretation.
Suppose that we control the system by using two time-dependent periodic
parameters $\bm{k}(t)=(k_1(t),k_2(t))$.
The adiabatic current $J_{\rm ad}$ arises only
when the orbit of $\bm{k}$ encloses a finite area.
The adiabatic current is represented by a flux penetrating the surface.
This geometrical picture is also applied to the nonadiabatic part.
We extend the parameter space and introduce
a third axis $k_3=\delta_2$.
Although $\delta_2$ is a function of $k_1$ and $k_2$,
we leave it independent for the moment and use
the relation after the calculation.
In the extended space $\tilde{\bm{k}}=(\bm{k},k_3)$,
$J_{\rm g}$ is written as 
\be
 J_{\rm g} = \oint_{\tilde{C}}\diff\tilde{\bm{k}}\cdot\bm{A}(\bm{k})
 = \int_{\tilde{S}}\diff\tilde{\bm{S}}(\tilde{\bm{k}})\cdot\bm{B}(\bm{k}),
 \label{stokes}
\ee 
where
$\tilde{C}$ represents the closed contour in the $\tilde{\bm{k}}$ space and 
$\bm{A}(\bm{k})$ is the ``gauge field'':
\be
 \bm{A}(\bm{k})=
 \frac{\omega}{2\pi}
 \bmat{c} p^{\rm (R)}\partial_{1}p_{\rm out} \\ 
 p^{\rm (R)}\partial_{2}p_{\rm out} \\ p^{\rm (R)} \emat.
\ee
This vector function is independent of $k_3$.
The adiabatic part is represented by the first and second components of $\bm{A}$
and the nonadiabatic part is by the third component.
We can introduce the corresponding ``magnetic field''
$\bm{B}(\bm{k}) = \bm{\nabla}\times\bm{A}(\bm{k})$.
The third (first and second) component of $\bm{B}(\bm{k})$ corresponds to  
the adiabatic (nonadiabatic) part. 
Using the Stokes theorem, we obtain the last expression in Eq.~(\ref{stokes}).
The integral represents a surface integral where the surface 
$\tilde{S}$ is defined by using the closed contour $\tilde{C}$.
This is pictorially represented as in Fig.~\ref{fig1}.
This surface is not unique and we can consider a convenient choice.
This geometrical representation does not mean that 
the result is independent of the control speed.
$\bm{B}$ is written in terms of purely geometric variables $k_1$ and $k_2$,
but the third axis is determined by the dynamics.

{\it Structure of the transition-rate matrix}.
Since the current is linear in $W$,
the decomposition of the current can also be applied to
the transition-rate matrix as $W(t)=W_{\rm d}(t)+W_{\rm g}(t)$.
The explicit form of $W_{\rm g}(t)$ is given by 
\be
 W_{\rm g}(t) = \left(\dot{p}_{\rm out}(t)+\dot{\delta}_2(t)\right)
 \bmat{cc} 1 & 1 \\ -1 & -1 \emat. \label{Wg}
\ee
The solution of the master equation
$|p(t)\rangle$ is given by the adiabatic state of $W_{\rm d}(t)$.
$W_{\rm g}(t)$ is interpreted as a counterdiabatic term 
known in shortcuts to adiabaticity 
(See SM)~\cite{DR03,DR05,Berry09,CRSCGM,STA13,STA19}.
It has a geometrical meaning~\cite{NT18},
which is consistent with the geometrical interpretation for $J_{\rm g}$. 

Using the decomposition of $W(t)$,
we can also find a relation to the Floquet theory.
The time-evolution operator
$U(t)={\rm T}\exp\left(\int_0^{t} \diff t'\, W(t')\right)$,
where ${\rm T}$ is the time-ordering operator, is
written at $t=T_0=2\pi/\omega$ as
$U(T_0)=\e^{T_0W_{\rm F}}$ to define
the effective transition-rate matrix $W_{\rm F}$.
Since the solution of the master equation is characterized
as a stationary state of $W_{\rm F}$,
$W_{\rm F}$ must be related to $W_{\rm d}(t)$.
In fact, we can write 
\be
 W_{\rm F}=\frac{\bar{k}}{k_{\rm in}(0)+k_{\rm out}(0)}
 \frac{W_{\rm d}(T_0)-\e^{-2\pi\bar{k}/\omega}W_{\rm d}(0)}
 {1-\e^{-2\pi\bar{k}/\omega}}, 
\ee
where $\bar{k}=\frac{1}{T_0}\int_0^{T_0}\diff t\,
\left(k_{\rm in}(t)+k_{\rm out}(t)\right)$.
See SM for details.
In the adiabatic limit $\omega\to 0$,
we find $W_{\rm F}\sim W_{\rm d}(T_0)$, which is consistent with the above
consideration.
In the opposite limit, the function
can be expanded in powers of $1/\omega$,
which is equivalent to the Floquet--Magnus expansion~\cite{Magnus,BCOR}.
This relation is useful since
we can find the decomposition of $W(t)$
by using the expansion at high frequencies.

\begin{center}
\begin{figure}[t]
\begin{center}
\includegraphics[width=1.\columnwidth]{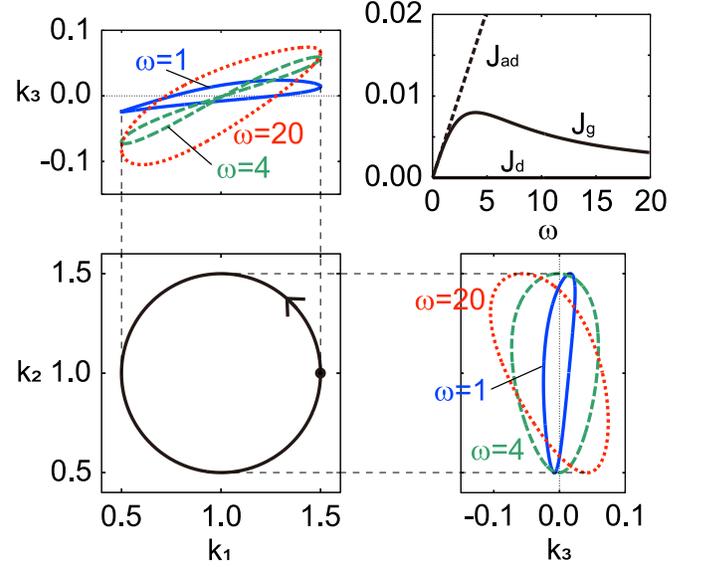}
\end{center}
\caption{
The frequency dependence of the current (top right) and 
trajectories in the parameter space at several values of $\omega$.
We set 
$k_1=k_{\rm in}^{\rm (L)}(t)=k_0\left(1+\frac{1}{2}\cos\omega t\right)$,
$k_2=k_{\rm in}^{\rm (R)}(t)=k_0\left(1+\frac{1}{2}\sin\omega t\right)$, 
$k_{\rm out}^{\rm (L)}=k_0$, and $k_{\rm out}^{\rm (R)}=k_0$.
All the quantities are plotted in unit of $k_0$.
}

\label{fig2}
\end{figure}
\end{center}

{\it Nonadiabatic effects on geometric current}.
A typical behavior of the current is shown in Fig.~\ref{fig2}.
We use a similar protocol as used in Ref.~\cite{SN07-1}.
Since we use a protocol with no net bias, the dynamical current is
negligibly small.
At low $\omega$, the adiabatic current is dominant,
which is proportional to $\omega$. 
It is considerably disturbed by the nonadiabatic effects at high $\omega$.
The total current approaches zero as $1/\omega$,
as is found from the Floquet--Magnus expansion.
Thus, the nonadiabatic effect inhibits 
the linearity of the geometric current with respect to $\omega$.

The behavior of the current is understood from the geometrical picture.
Since the third component of the flux 
$B_3(\bm{k})=\partial_1A_2(\bm{k})-\partial_2A_1(\bm{k})$ 
determines the adiabatic current, 
the geometric current coincides with the adiabatic current if
the trajectory $\tilde{\rm C}$ is parallel to the $(k_1, k_2)$ plane.
In Fig.~\ref{fig2}, we see that, as the frequency increases, 
the trajectory is distorted from a flat plane
to cancel out the adiabatic part.

\begin{center}
\begin{figure}[t]
\begin{center}
\includegraphics[width=1.\columnwidth]{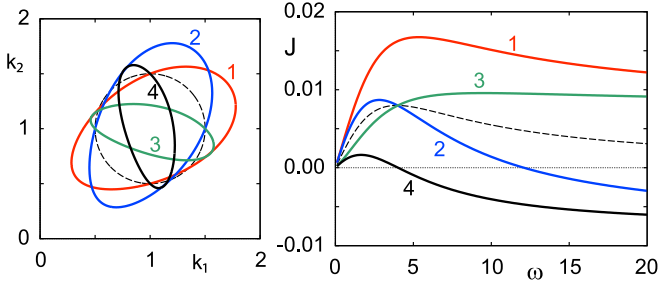}
\end{center}
\caption{
Left: Elliptic trajectories (1--4)
keeping the dynamical current invariant.
The dashed line represents the original protocol used in Fig.~\ref{fig2}.
Right: The corresponding total current. 
The dynamical part is zero in each protocol.
See SM for details.}
\label{fig3}
\end{figure}
\end{center}

In Fig.~\ref{fig3}, we plot the current when
the trajectory ${\rm C}$ is slightly deformed
while keeping the dynamical current invariant (See SM for details).
We still observe nonadiabatic effects affecting linear growth.
To keep the adiabatic current, 
we need to design the protocol such that the plane is kept
parallel to the $(k_1, k_2)$ plane.
Since we cannot choose the trajectory $\tilde{\rm C}$ arbitrary, 
this is a difficult problem in general.

{\it Assisted adiabatic pumping}.
To obtain a desirable enhancement of the geometric current, 
we use a method of counterdiabatic driving.
We introduce the counterdiabatic term into the original transition matrix
so that the adiabatic state of the original matrix
becomes the exact solution.
Although the idea is implemented for the Schr\"odinger equation
for isolated quantum systems,
the generalization to other equations such as the master equation
and the Fokker--Planck equation is a straightforward task.
We can find several applications in previous studies~\cite{IMCTM,Tu,TO,LQT}.

In the master equation, the transition-rate matrix is diagonalized as 
$W(t)=\sum_n\epsilon_n(t)|\phi_n(t)\rangle\langle\phi_n(t)|$ and 
the adiabatic state is defined by Eq.~(\ref{p0})
with time-independent coefficients $\{C_n\}$.
We modify the transition-rate matrix $W(t) \to W(t)+W_{\rm CD}(t)$
so that the solution of the modified master equation 
is given by the adiabatic state.
The counterdiabatic term $W_{\rm CD}(t)$ is given by 
\be
 W_{\rm CD}(t)
 = \sum_{m,n(m\ne n)}|\phi_m(t)\rangle\langle \phi_m(t)|\dot{\phi}_n(t)\rangle
 \langle \phi_n(t)|.
\ee
For the two-state case, $W_{\rm CD}(t)$ can be explicitly written as 
\be
  W_{\rm CD}(t)= \dot{p}_{\rm out}(t)\bmat{cc} 1 & 1 \\ -1 & -1 \emat. 
\ee
This form is slightly different from $W_{\rm g}(t)$ in Eq.~(\ref{Wg}).
We see that the addition of the counterdiabatic term 
is obtained by replacements 
$k_{\rm in}(t)\to k_{\rm in}(t)-\dot{p}_{\rm out}(t)$ and 
$k_{\rm out}(t)\to k_{\rm out}(t)+\dot{p}_{\rm out}(t)$.
Since these variables represent the transition rates,
$|\dot{p}_{\rm out}(t)|$ cannot be large and
the method fails for rapid changes of parameters.

The inclusion of the counterdiabatic term ensures that
the exact solution of the master equation is given by the adiabatic state
of the original transition-rate matrix.
$k_{\rm in}(t)$ and $k_{\rm out}(t)$ are, respectively, represented by 
the sum of the left and right parts and we still have
degrees of freedom to implement the counterdiabatic term.
We can use them to keep the dynamical part of the current invariant
and to set that the geometric part of the current is given by 
the adiabatic part of the original current without assist (See SM).

\begin{center}
\begin{figure}[t]
\begin{center}
\includegraphics[width=1.\columnwidth]{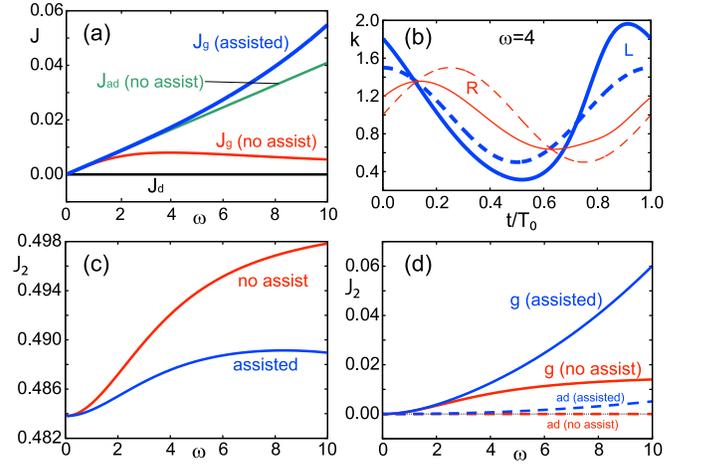}
\end{center}
\caption{
Assisted adiabatic pumping.
We use the same protocol as used in Fig.~\ref{fig2}
for the original system before assist.  
(a) The geometric current with an assisting field is 
represented by the blue line.
The other lines are the same as in the top right panel of Fig.~\ref{fig2}.
(b) Dashed lines representing protocols before assist 
are changed to solid lines by the assisting field.
Bold blue lines represent the left amplitude $k_{\rm in}^{\rm (L)}$
and thin red lines do the right amplitude $k_{\rm in}^{\rm (R)}$.
We take $\omega=4.0$.
(c) The current fluctuations before/after assist (See SM).
(d) The geometric part (solid lines) and 
the adiabatic part (dashed lines) of the current fluctuations.
}
\label{fig4}
\end{figure}
\end{center}

Although the above procedure works in principle, 
we have no clear picture on how the assisting field enhances the current.
In addition, the manipulation is restricted in realistic situations and
we cannot control each element in the transition-rate matrix independently. 
In our choice in the above examples, we set that 
$k_{\rm out}$ is time independent.
The introduction of the counterdiabatic term inevitably breaks this condition. 
To keep the time independence of $k_{\rm out}$, we can consider scaling.
After the introduction of the counterdiabatic term, 
we write the transition-rate matrix as 
\be
 W(t)+W_{\rm CD}(t)  = 
 \left(1+\frac{\dot{p}_{\rm out}(t)}{k_{\rm out}}\right)\bmat{cc}
 {\displaystyle
 -\frac{1-\frac{\dot{p}_{\rm out}(t)}{k_{\rm in}(t)}}{1+\frac{\dot{p}_{\rm out}(t)}{k_{\rm out}}}
 k_{\rm in}(t)} & k_{\rm out} \\
 {\displaystyle \frac{1-\frac{\dot{p}_{\rm out}(t)}{k_{\rm in}(t)}}{1+\frac{\dot{p}_{\rm out}(t)}{k_{\rm out}}}k_{\rm in}(t)} & -k_{\rm out} \emat. \no\\
\ee
The prefactor of the right-hand side is positive and is scaled out by 
the redefinition of the timescale as 
$\diff\tilde{t}=\diff t (1+\dot{p}_{\rm out}(t)/k_{\rm out})$.
We still have a degree of freedom to decompose 
the new component $k_{\rm in}(t)$ into the left and right parts
and use it to keep the dynamical current invariant.
In this case, the geometric current is not equal 
to the adiabatic current in the original system
and is not proportional to the frequency.
However, we confirm that the deviation is not so large 
and the geometric current can be kept growing as a function of the frequency.
The result is shown in Fig.~\ref{fig4} (See SM for details).
The obtained protocol indicates that we need to shift the oscillation 
of the assisting field to the left compared to the original one 
to prevent the deviation.
The required field becomes larger when we consider faster driving and
the control fails at some frequency where $|\dot{p}_{\rm out}(t)|$
exceeds the threshold.

In Fig.~\ref{fig4}, we also plot the current fluctuation that 
is decreasing by the introduction of the assisting field.
Generally, the counterdiabatic term leads to
an increase in the energy cost characterized by the fluctuation and 
a broadening of the work distribution~\cite{CD,FZCKUC}.
This expectation i.e. the increment of the fluctuation for 
the geometric part under the assisting field 
is verified as can be seen
on the bottom right panel of Fig.~\ref{fig4}.
Although we cannot control the dynamical part of the fluctuation 
as we did for the average, we find a decrease of the total 
fluctuation as a result of the decrease of the dynamical fluctuation.
The suppression of the fluctuations implies the stability 
of the assisted driving.
A variational formulation of the counterdiabatic driving
for quantum systems also indicates a stable driving~\cite{Takahashi}.
In SM, we examine several examples to confirm the stability 
by slightly modifying the amplitudes in several ways.

{\it Note added in proof}.
After the completion of this work,
we learned about Ref.~\cite{FLNF} where a similar method is used
for adiabatic pumping.

We are grateful to Adolfo del Campo, Ken Funo, Jun Ohkubo, and Keiji Saito 
for useful discussions and comments.
We also thank Ville Paasonen for his critical reading of this manuscript.
This work was supported by JSPS KAKENHI Grant 
No. JP19J13698 (K.~F.), No. JP16H04025 (H.~H. and Y.~H.), 
and No. JP26400385 (K.~T.). 
K.~T. acknowledges the warm hospitality of the Yukawa Institute for 
Theoretical Physics, Kyoto University during his stay there 
to promote the collaboration among the authors. 


\onecolumngrid

\def\theequation{S\arabic{equation}}
\makeatletter
\@addtoreset{equation}{section}
\makeatother

\setcounter{equation}{0}

\newpage
\section{Supplemental Material}

\twocolumngrid

\section{Master equation}

\subsection{Improved adiabatic approximation}

We want to solve the master equation
\be
 \frac{\diff}{\diff t}|p(t)\rangle = W(t)|p(t)\rangle.
\ee
We assume that the matrix is diagonalizable.
Then, the instantaneous eigenstates of $W(t)$ are prepared as 
\be
 && W(t)|\phi_n(t)\rangle = \epsilon_n(t)|\phi_n(t)\rangle, \\
 && \langle \phi_n(t)|W(t) = \langle\phi_n(t)|\epsilon_n(t).
\ee
We have the orthonormal relations and the resolution of unity: 
\be
 && \langle \phi_m(t)|\phi_n(t)\rangle = \delta_{m,n}, \label{on}\\
 &&  \sum_n|\phi_n(t)\rangle\langle \phi_n(t)| = 1. \label{c}
\ee
The left and the right eigenstates are not a simple conjugate with each other.
We also assume that $n=1$ represents the stationary state 
and the other states represent decaying contributions, which means 
that the eigenvalues satisfy 
\be
 && \epsilon_1(t) = 0, \\
 && \epsilon_n(t) < 0 \quad(n\ne 1).
\ee
Although it is difficult to find a specific form of the eigenstates in general, 
$\langle\phi_1(t)|$ has a simple form as 
\be
 \langle\phi_1(t)| 
 = \bmat{cccc} 1 & 1 & \cdots & 1 \emat =:\langle 1|, \label{bra1}
\ee
due to the property of the transition-rate matrix $\sum_i W_{ij}(t)=0$.

The eigenstates have degrees of freedom as 
\be
 && |\phi_n(t)\rangle \to R_n(t)|\phi_n(t)\rangle, \\
 && \langle \phi_n(t)| \to \langle\phi_n(t)|R_n^{-1}(t),
\ee
where $R_n(t)$ is an arbitrary function with $R_n(0)=1$.
To remove this arbitrariness, we introduce 
\be
 && |\tilde{\phi}_n(t)\rangle 
 = \e^{-\int_0^t\diff t'\,\langle \phi_n(t')|\dot{\phi}_n(t')\rangle}|\phi_n(t)\rangle, \\
 && \langle\tilde{\phi}_n(t)|
 = \langle\phi_n(t)|\e^{\int_0^t\diff t'\,\langle \phi_n(t')|\dot{\phi}_n(t')\rangle}.
\ee
These eigenstates are invariant under the transformation of $R_n(t)$.
We note that the transformation does not change the properties in 
Eqs.~(\ref{on}) and (\ref{c}).
We also have for any $n$
\be
 \langle \tilde{\phi}_n(t)|\dot{\tilde{\phi}}_n(t)\rangle = 0.
 \label{ndn}
\ee

We expand the solution of the master equation with respect to 
$\{|\tilde{\phi}_n(t)\rangle\}$ as 
\be
 |p(t)\rangle = \sum_n C_n(t)\e^{\int_0^t \diff t'\,\epsilon_n(t')}
 |\tilde{\phi}_n(t)\rangle. \label{psum}
\ee
$C_1(t)$ is determined from the normalization as 
\be
 C_1(t) = \langle \tilde{\phi}_1(t)|p(t)\rangle
 = \langle 1|p(t)\rangle = 1.
\ee
To solve the other components, we substitute the representation (\ref{psum})
to the master equation and multiply $\langle\tilde{\phi}_n(t)|$ 
from the left.
We obtain 
\be
 && \sum_{m(\ne 1)} C_m(t)\e^{\int_0^t\diff t'\,\epsilon_m(t') }
 \langle \tilde{\phi}_1(t)|\dot{\tilde{\phi}}_m(t)\rangle = 0, \label{eq0}\\
 &&\frac{\diff C_n(t)}{\diff t}\e^{\int_0^t\diff t'\,\epsilon_n(t') }
 +\langle \tilde{\phi}_n(t)|\dot{\tilde{\phi}}_1(t)\rangle \no\\
 && +\sum_{m(\ne 1,n)} C_m(t)\e^{\int_0^t\diff t'\,\epsilon_m(t') }
 \langle \tilde{\phi}_n(t)|\dot{\tilde{\phi}}_m(t)\rangle = 0.
 \label{eq}
\ee
In the second equation, 
the contribution of $m=1$ is separated from the sum.

As we mention in the main body of the paper, 
we neglect the third term on the left hand side of Eq.~(\ref{eq}). 
Then, we obtain
\be
 C_n(t)\simeq C_n(0)-\int_0^t \diff t'\, 
\langle \tilde{\phi}_n(t')|\dot{\tilde{\phi}}_1(t')\rangle
\e^{-\int_0^{t'}\diff t''\,\epsilon_n(t'') }.
\ee
The solution of the master equation is approximated to
\be
 |p(t)\rangle \simeq |\tilde{\phi}_1(t)\rangle
 +\sum_{n(\ne 1)}\left(\delta_n(t)+C_n(0)\e^{\int_{0}^{t} \diff t'\,\epsilon_n(t')}\right)
 |\tilde{\phi}_n(t)\rangle, \no\\
\ee
where 
\be
 \delta_n(t) = -\int_0^t \diff t'\, 
 \langle \tilde{\phi}_n(t')|\dot{\tilde{\phi}}_1(t')\rangle 
 \e^{\int_{t'}^{t}\diff t''\,\epsilon_n(t'') }. \label{deltan}
\ee

\subsection{Exact solution for two-state system}

To obtain an explicit form of the state,
we examine the two-state case.
The transition-rate matrix is generally written as 
\be
 W(t)=\bmat{cc}
 -k_{\rm in}(t) & k_{\rm out}(t) \\
 k_{\rm in}(t) & -k_{\rm out}(t)
 \emat,
\ee
where $k_{\rm in}(t)$ and $k_{\rm out}(t)$ are arbitrary nonnegative functions.
The instantaneous eigenstates of $W(t)$ are given by  
\be
 && \{|\phi_n(t)\rangle\}_{n=1,2} = \left\{
 \bmat{c} p_{\rm out}(t) \\ 1-p_{\rm out}(t) \emat, \quad
 \bmat{c} 1 \\ -1 \emat
 \right\}, \\
 && \{\langle \phi_n(t)|\}_{n=1,2} =\left\{
 \bmat{cc} 1 & 1 \emat, \quad
 \bmat{cc} 1-p_{\rm out}(t) & -p_{\rm out}(t) \emat
 \right\}, \no\\ 
\ee
where 
\be
 p_{\rm out}(t)=\frac{k_{\rm out}(t)}{k_{\rm in}(t)+k_{\rm out}(t)}.
\ee
The corresponding eigenvalues are 
$\{\epsilon_n(t)\}_{n=1,2}=\{0,-(k_{\rm in}(t)+k_{\rm out}(t))\}$.
The component $n=1$ represents the instantaneous stationary state.
In this case, the geometric phase is zero in each level and 
we have $|\tilde{\phi}_n(t)\rangle=|\phi_n(t)\rangle$ and
$\langle\tilde{\phi}_n(t)|=\langle\phi_n(t)|$.

Now we expand the solution as in Eq.~(\ref{psum}).
Using the master equation, we obtain
\be
 && \frac{\diff C_1(t)}{\diff t}= 0, \\
 && \frac{\diff C_2(t)}{\diff t}=-C_1(t)\e^{\int_0^t \diff t'\,(\epsilon_1(t')-\epsilon_2(t'))}\langle \phi_2(t)|\dot{\phi}_1(t)\rangle.
\ee
The first equation shows that $C_1$ is independent of $t$, and 
the second equation can be solved simply by integrating the equation.
With the initial condition $|p(0)\rangle = (p_0, 1-p_0)^{\rm T}$, 
we obtain the exact result:
\be
 |p(t)\rangle 
 &=& \bmat{c} p_{\rm out}(t)+\delta(t) \\ 1-p_{\rm out}(t)-\delta(t) \emat \no\\
 && + \left(p_0-p_{\rm out}(0)\right)\e^{-\int_0^t \diff t'\,(k_{\rm in}(t')+k_{\rm out}(t'))}\bmat{c} 1 \\ -1 \emat, \label{pt}
\ee
where 
\be
 \delta(t) = -\int_0^t \diff t'\,\dot{p}_{\rm out}(t')
 \e^{-\int_{t'}^{t} \diff t''\,(k_{\rm in}(t'')+k_{\rm out}(t''))}. \label{delta}
\ee
$\delta(t)$ is equivalent to $\delta_n(t)$ in Eq.~(\ref{deltan}) with $n=2$.
The dependence of the initial condition is only 
in the last term of Eq.~(\ref{pt}).
This term decays exponentially as a function of $t$.
Combining with the property of $\delta(t)$ discussed below, 
we can conclude that the system rapidly approaches 
a periodic behavior which is
independent of the initial condition
and the pumping current is independent of the second term of Eq.~(\ref{pt}).

Generally, the time evolution operator between two states, 
defined by $|p(t_2)\rangle=U(t_2,t_1)|p(t_1)\rangle$, is given by 
\be
 &&U(t_2,t_1)=|\phi_1(t_2)\rangle\langle 1|
 +\delta(t_2)|2\rangle\langle 1| \no\\
 && +\e^{-\int_{t_1}^{t_2}\diff t\,(k_{\rm in}(t)+k_{\rm out}(t))}
 \left(|2\rangle\langle\phi_2(t_1)|-\delta(t_1)|2\rangle\langle 1|\right),
\ee
where $|2\rangle=|\phi_2(t)\rangle$ is independent of $t$. 
This operator can be written in a matrix form as 
\be
 U(t_2,t_1) &=&1+\frac{W_{\rm d}(t_2)}{k_{\rm in}(t_2)+k_{\rm out}(t_2)} \no\\
 &&-\e^{-\int_{t_1}^{t_2}\diff t\,(k_{\rm in}(t)+k_{\rm out}(t))}
 \frac{W_{\rm d}(t_1)}{k_{\rm in}(t_1)+k_{\rm out}(t_1)}, \label{U}
\ee
where
\be
 \frac{W_{\rm d}(t)}{k_{\rm in}(t)+k_{\rm out}(t)} =
 \bmat{cc} -(1-p_{\rm out}(t)-\delta(t)) & p_{\rm out}(t)+\delta(t) \\
 1-p_{\rm out}(t)-\delta(t) & -(p_{\rm out}(t)+\delta(t)) \emat. \no\\
 \label{Wd}
\ee
In the following calculations, we use the time-evolution operator
to obtain the current fluctuations and
the Floquet transition-rate matrix. 

\section{On the behavior of $\delta(t)$}

The nonadiabatic effects are determined by
$\delta_n(t)$ in Eq.~(\ref{deltan}).
Since the structure of the function is unchanged for any choice of $n$, 
we study the two-state case with $n=2$.
$\delta(t)=\delta_2(t)$ defined in Eq.~(\ref{delta}) 
satisfies the differential equation
\be
 \frac{\diff\delta(t)}{\diff t}=-(k_{\rm in}(t)+k_{\rm out}(t))
 \left(\delta(t)+\frac{\dot{p}_{\rm out}(t)}{k_{\rm in}(t)+k_{\rm out}(t)}\right). 
 \label{dotdelta}
\ee
We see that $\delta(t)=\delta^{(0)}(t)$ with 
$\delta^{(0)}(t):=-\dot{p}_{\rm out}(t)/(k_{\rm in}(t)+k_{\rm out}(t))$ 
represents the stationary point.
This point is stable against the deviation.
Therefore, if $p_{\rm out}(t)$ changes very slowly, $\delta(t)\simeq\delta^{(0)}(t)$
becomes a good approximation.

To improve the approximation, we consider the derivative expansion.
Equation (\ref{dotdelta}) is rewritten as 
\be
 \delta(t) = \delta^{(0)}(t) -\frac{1}{k_{\rm in}(t)+k_{\rm out}(t)}\frac{\diff}{\diff t}\delta(t). \label{delta2}
\ee
Solving the equation recursively, we obtain 
\be
 \delta(t) &=& \delta^{(0)}(t)
 +\left(-\frac{1}{k_{\rm in}(t)+k_{\rm out}(t)}\frac{\diff}{\diff t}\right)\delta^{(0)}(t) \no\\
 && +\left(-\frac{1}{k_{\rm in}(t)+k_{\rm out}(t)}\frac{\diff}{\diff t}\right)^2\delta^{(0)}(t)+\cdots. \label{smallom}
\ee
When each parameter is written as a function of $\omega t$, 
this is a series expansion of $\omega$ 
for a fixed $\omega t$.
The first term is the first order in $\omega$, 
the second term is the second order, and so on.

We plot $\delta(t)$ in Fig.~\ref{fig:delta}.
We consider the following periodic driving: 
\be
 && k_{\rm in}^{\rm (L)}(t)=k_0\left(1+\frac{1}{2}\cos\omega t\right), \\
 && k_{\rm in}^{\rm (R)}(t)=k_0\left(1+\frac{1}{2}\sin\omega t\right),  \\
 && k_{\rm out}^{\rm (L)}(t)=k_0, \\
 && k_{\rm out}^{\rm (R)}(t)=k_0. 
\ee
$k_0$ represents a constant.
As we see in the figure, $\delta(t)$ is almost periodic in $t$ 
for any choice of parameters.
It can be approximated to the stationary value 
$\delta^{(0)}(t)$ at small-$\omega$.
The deviation is described by the expansion in Eq.~(\ref{smallom}).

In the opposite limit where $\omega$ is large, 
$\delta(t)$ is approximated to $-(p_{\rm out}(t)-p_{\rm out}(0))$.
This is obtained by neglecting $\delta(t)$ 
in the right hand side of Eq.~(\ref{dotdelta}).
The $1/\omega$-correction can be evaluated by using 
the Floquet--Magnus expansion.

\begin{widetext}
\begin{center}
\begin{figure}[t]
\begin{center}
\begin{minipage}[h]{0.45\textwidth}
\begin{center}
\includegraphics[width=1.\columnwidth]{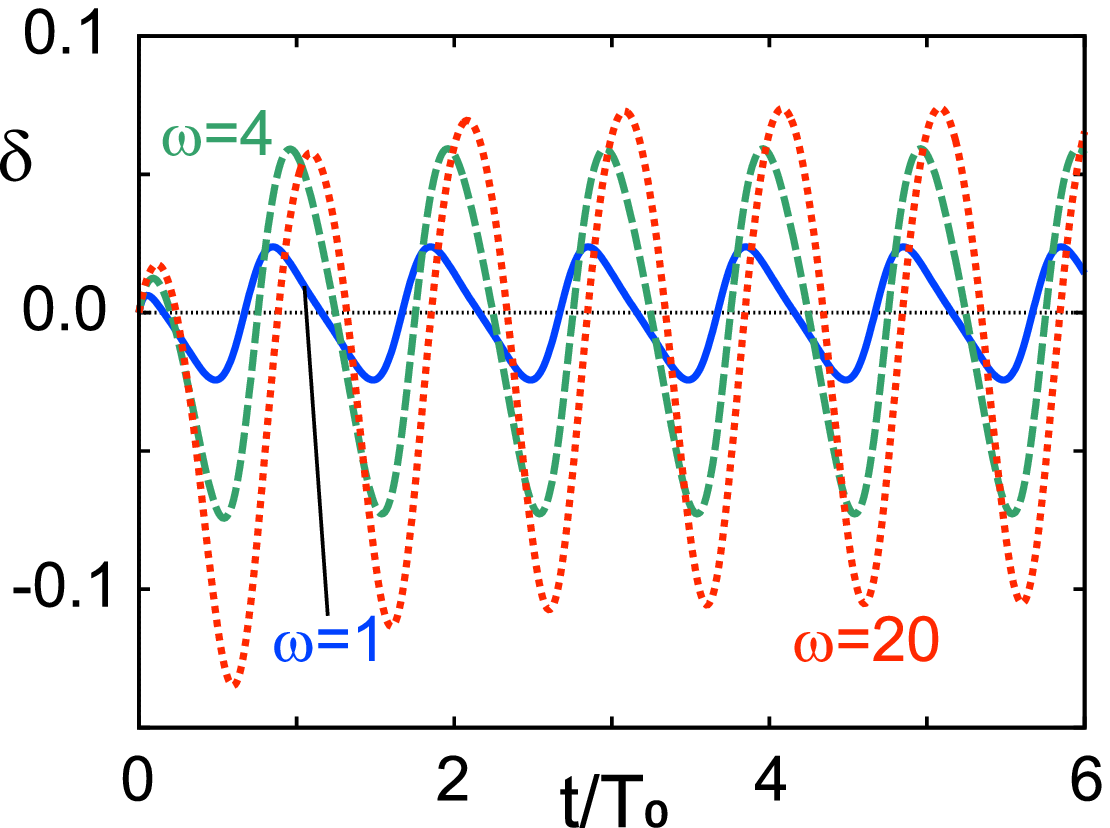}
\end{center}\end{minipage}\hspace*{0.1\textwidth}
\begin{minipage}[h]{0.45\textwidth}
\begin{center}
\end{center}\end{minipage}
\begin{minipage}[h]{0.45\textwidth}
\begin{center}
\includegraphics[width=1.\columnwidth]{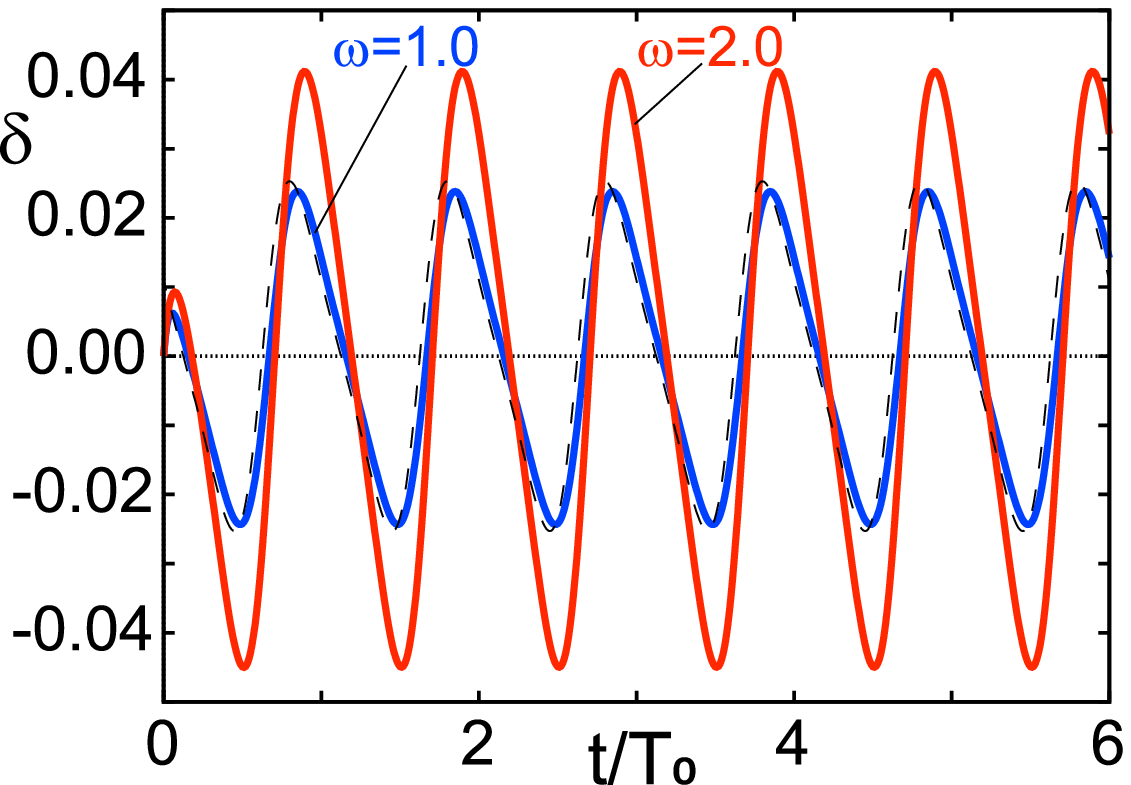}
\end{center}\end{minipage}\hspace*{0.1\textwidth}
\begin{minipage}[h]{0.45\textwidth}
\begin{center}
\includegraphics[width=1.\columnwidth]{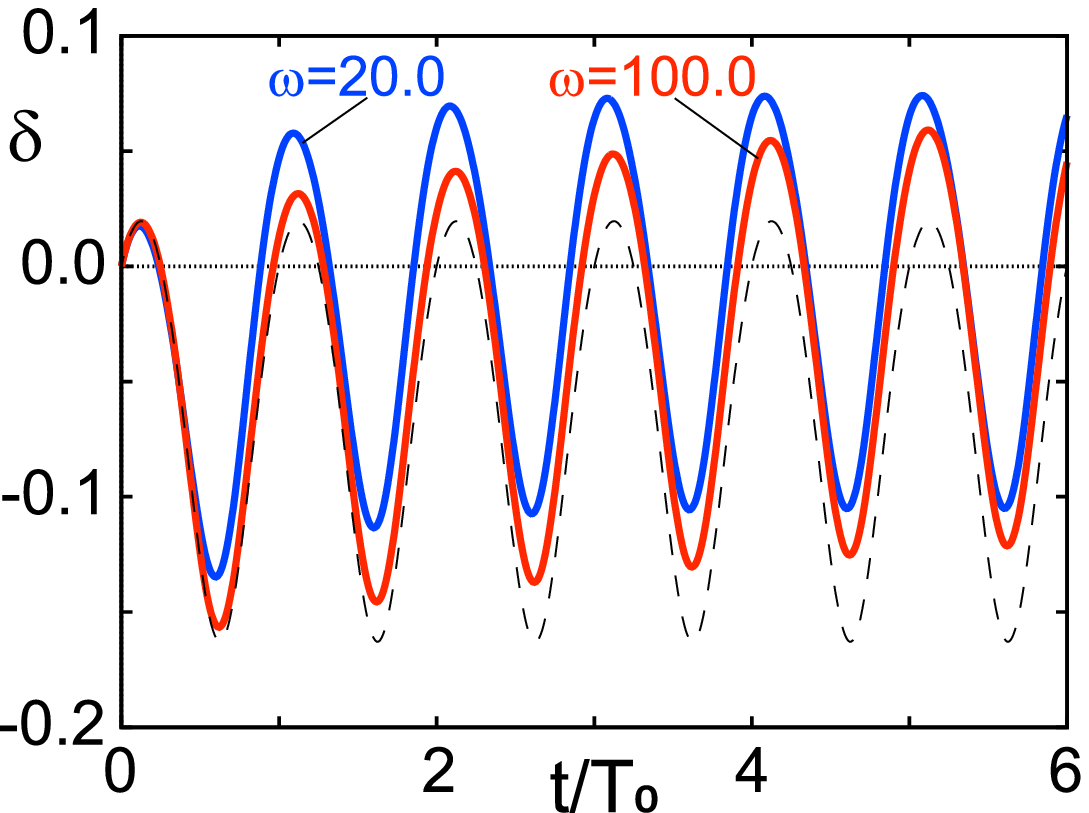}
\end{center}\end{minipage}
\end{center}
\caption{Top: $\delta(t)$ for several values of $\omega$.
Bottom Left: $\delta(t)$ for a slow driving (small $\omega$). 
$\delta(t)$ at $\omega=1.0$ is denoted by the blue solid line 
and $\omega=2.0$ by the red solid line.
The black dashed line denotes the stationary point 
$\delta^{(0)}(t)=-\dot{p}_{\rm out}(t)/(k_{\rm in}(t)+k_{\rm out}(t))$.
Bottom Right: $\delta(t)$ for a fast driving (large $\omega$).
$\delta(t)$ at $\omega=20.0$ is denoted by the blue solid line 
and $\omega=100.0$ by the red solid line.
The black dashed line denotes the asymptotic result
$\delta(t)\simeq -p_{\rm out}(t)+p_{\rm out}(0)$.
}
\label{fig:delta}
\end{figure}
\end{center}
\end{widetext}

\section{Counting field and current distributions}
\label{cf}

\subsection{Counting field}

The current distribution function is calculated by introducing the counting
field $\chi$ to the transition-rate matrix as $W(t)\to W(t;\chi)$.
The explicit form is given by 
\be
 W(t;\chi)=\bmat{cc}
 -k_{\rm in}(t) & k_{\rm out}^{\rm (L)}(t)+k_{\rm out}^{\rm (R)}(t)\e^{i\chi} \\
 k_{\rm in}^{\rm (L)}(t)+k_{\rm in}^{\rm (R)}(t)\e^{-i\chi} & -k_{\rm out}(t)
 \emat. \no\\
\ee
Using the solution of the master equation $|p(t;\chi)\rangle$ with
the modified matrix $W(t;\chi)$, we write 
\be
 \langle 1|p(t;\chi)\rangle =\exp\left(i\chi n(t)-\frac{\chi^2}{2}n_2(t)+\cdots\right),
\ee
and the average current and the fluctuation is given by
\be
 && J = \lim_{T\to\infty}\frac{1}{T}n(T), \\
 && J_2 = \lim_{T\to\infty}\frac{1}{T}n_2(T).
\ee
We note that $J_2$ represents the second-order cumulant 
$\langle\hat{J}^2\rangle-\langle\hat{J}\rangle^2$.

To calculate the current distributions,
we expand the matrix $W(t;\chi)$ as 
\be
 && W(t;\chi)=W(t)+i\chi V_1(t)-\frac{\chi^2}{2}V_2(t)+\cdots, \\
 && V_1(t) = \bmat{cc}
 0 & k_{\rm out}^{\rm (R)}(t) \\
 -k_{\rm in}^{\rm (R)}(t) & 0
 \emat, \\
 && V_2(t) = \bmat{cc}
 0 & k_{\rm out}^{\rm (R)}(t) \\
 k_{\rm in}^{\rm (R)}(t) & 0
 \emat.
\ee

\subsection{Current distributions}

Using the derived formula, we find that the average current is given by  
\be
 J &=& \lim_{T\to\infty}\frac{1}{T}\int_0^T \diff t\, \langle 1|V_1(t)|p(t)\rangle
 \no\\
 &=& \lim_{T\to\infty}\frac{1}{T}\int_0^T \diff t\, 
 \left(k_{\rm out}^{\rm (R)}(t)p_2(t)-k_{\rm in}^{\rm (R)}(t)p_1(t)\right).
\ee
Using Eq.~(\ref{pt}), we have 
\be
 J &=& \lim_{T\to\infty}\frac{1}{T}\int_0^T \diff t\,
 \Bigl[
 k_{\rm out}^{\rm (R)}(t)(1-p_{\rm out}(t))-k_{\rm in}^{\rm (R)}(t)p_{\rm out}(t) \no\\
&& -\left(k_{\rm out}^{\rm (R)}(t)+k_{\rm in}^{\rm (R)}(t)\right)\delta(t)
  \Bigr].
\ee
Since the second term of Eq.~(\ref{pt}) incorporates an exponential factor, 
it does not contribute to the result.
Then, we find that the current is independent on the initial condition.
The dynamical part of the current is given by setting
$\delta(t)=0$.
The decomposition of the geometric part  into the adiabatic part
and the nonadiabatic part can be found by using Eq.~(\ref{delta2}).
The explicit form of each part is given in the main body of the paper.

Similarly, the fluctuation is obtained from
\be
 &&-\frac{1}{2}\left(n_2(t)+n^2(t)\right) 
 = -\frac{1}{2}\int_0^t \diff t'\,\langle 1|V_2(t')|p(t')\rangle \no\\
 && -\int_0^t \diff t_2\int_0^{t_2} \diff t_1\,
 \langle 1|V_1(t_2)U(t_2,t_1)V_1(t_1)|p(t_1)\rangle.
\ee
After some calculations, we obtain
\be
 &&J_2 = \lim_{T\to\infty}\frac{1}{T}\int_0^T \diff t\,
 \Bigl[
 k_{\rm out}^{\rm (R)}(t)p_{\rm in}(t)+k_{\rm in}^{\rm (R)}(t)p_{\rm out}(t) \no\\
 &&-\left(k_{\rm out}^{\rm (R)}(t)-k_{\rm in}^{\rm (R)}(t)\right)\delta(t) 
 +2\left(k_{\rm out}^{\rm (R)}(t)+k_{\rm in}^{\rm (R)}(t)\right)\Delta(t)
 \Bigr],
\ee
where 
\be
 &&\Delta(t)= -\int_0^t \diff t'\,\Bigl[
 k_{\rm out}^{\rm (R)}(t')(p_{\rm in}(t')-\delta(t'))^2 \no\\
 && +k_{\rm in}^{\rm (R)}(t')(p_{\rm out}(t')+\delta(t'))^2  
 \Bigr]\e^{-\int_{t'}^t \diff t''\,(k_{\rm in}(t'')+k_{\rm out}(t''))}.
\ee
$\Delta(t)$ satisfies a first-order differential equation which has a similar
form to that for $\delta(t)$ and
its behavior can also be understood in a similar way.
It is a straightforward task to decompose $J_2$ into 
dynamical, adiabatic, and nonadiabatic parts and
we do not show their explicit forms here.

\section{Decomposition of the transition-rate matrix}

In the time-evolution operator,
we have introduced a matrix $W_{\rm d}(t)$ in Eq.~(\ref{Wd}).
It satisfies the eigenvalue equations
\be
 && W_{\rm d}(t)\bmat{c} p_{\rm out}(t)+\delta(t) \\ 1-p_{\rm out}(t)-\delta(t) \emat
 = 0, \\
 && W_{\rm d}(t)\bmat{c} 1 \\ -1 \emat
 = -(k_{\rm in}(t)+k_{\rm out}(t))\bmat{c} 1 \\ -1 \emat.
\ee
Each vector appears in the solution of the master equation (\ref{pt}). 
This property shows that $|p(t)\rangle$ is equal to the adiabatic
state of $W_{\rm d}(t)$.
Then, $W(t)$ is decomposed as 
\be
 W(t)=W_{\rm d}(t)+W_{\rm g}(t),
\ee
and $W_{\rm g}(t)$ plays the role of the counterdiabatic term.
Using Eq.~(\ref{dotdelta}), we can write $W_{\rm g}(t)$ as 
\be
 W_{\rm g}(t)=(\dot{p}_{\rm out}(t)+\dot{\delta}(t))
 \bmat{cc} 1 & 1 \\ -1 & -1 \emat.
\ee
This term gives the geometric current as we see from the relation 
\be
 (W_{\rm g}(t))_{12}^{\rm (R)}p_2(t)- (W_{\rm g}(t))_{21}^{\rm (R)}p_1(t) 
 = p^{\rm (R)}(t)\left(\dot{p}_{\rm out}(t)+\dot{\delta}(t)\right). \no\\ 
\ee

\section{Floquet theory}

We derive the Floquet effective transition-rate matrix $W_{\rm F}$
defined as 
\be
 U(T_0,0)=\e^{T_0W_{\rm F}}.
\ee
The time-evolution operator is given in Eq.~(\ref{U}).
It has a form 
\be
 U(t_2,t_1)=1+X(t_2,t_1), 
\ee
and the matrix $X$ satisfies the relation 
\be
 X^2(t_2,t_1)=-\left(1-\e^{-\int_{t_1}^{t_2}\diff t\,(k_{\rm in}(t)+k_{\rm out}(t))}\right)
 X(t_2,t_1).
\ee
This is a very convenient formula to take the logarithm of $U$.
We easily find 
\be
 W_{\rm F}T_0 &=& \frac{\int_{0}^{T_0}\diff t\,(k_{\rm in}(t)+k_{\rm out}(t))}
 {1-\e^{-\int_{0}^{T_0}\diff t\,(k_{\rm in}(t)+k_{\rm out}(t))}}
 \frac{1}{k_{\rm in}(0)+k_{\rm out}(0)} \no\\
 && \times \left( W_{\rm d}(T_0)
 -\e^{-\int_{0}^{T_0}\diff t\,(k_{\rm in}(t)+k_{\rm out}(t))}W_{\rm d}(0)
 \right).
\ee
To make the frequency dependence clear, we
use
\be
 \bar{k}=\frac{1}{T_0}\int_0^{T_0}\diff t\,\left(k_{\rm in}(t)+k_{\rm out}(t)\right),
\ee
to write 
\be
 W_{\rm F} = \bar{k}
 \left[
 \bmat{cc} -p_{\rm in}(0) & p_{\rm out}(0) \\ p_{\rm in}(0) & -p_{\rm out}(0) \emat
 +\frac{\delta(T_0)}{1-\e^{-2\pi\bar{k}/\omega}}\bmat{cc} 1 & 1 \\ -1 & -1 \emat
 \right]. \no\\
\ee
We note that this is the exact form of the effective transition-rate matrix.
The expression can be expanded in powers of $1/\omega$.
We use the asymptotic expansion
$\delta(t)\sim -(p_{\rm out}(t)-p_{\rm out}(0))+\cdots$.
At the leading order, we have
\be
 W_{\rm F}\sim \frac{1}{T_0}\int_0^{T_0} \diff t\,W(t).
\ee
This is the zeroth order contribution in the Floquet--Magnus expansion,
though Hamiltonian in the conventional Floquet--Magnus theory
is replaced by the transition-rate matrix.
Similarly, we can find the higher-order contributions.

\section{Deformation of the protocol trajectory}

The local dynamical current is given by 
\be
 J_{\rm d}(t)=
 \frac{k_{\rm in}^{\rm (L)}(t)k_{\rm out}^{\rm (R)}(t)-k_{\rm out}^{\rm (L)}(t)k_{\rm in}^{\rm (R)}(t)}{k(t)}. \label{Jdt}
\ee
We can easily confirm that $J_{\rm d}(t)$ is invariant under 
the transformation
\be
 && k_{\rm in}^{\rm (L)}(t)\to k_{\rm in}^{\rm (L)}(t)+k^{\rm (L)}(t)f(t), \\
 && k_{\rm in}^{\rm (R)}(t)\to k_{\rm in}^{\rm (R)}(t)+k^{\rm (R)}(t)f(t),
\ee
where 
$k^{\rm (L)}=k_{\rm in}^{\rm (L)}+k_{\rm out}^{\rm (L)}$,  
$k^{\rm (R)}=k_{\rm in}^{\rm (R)}+k_{\rm out}^{\rm (R)}$, and
$f(t)$ is an arbitrary function.

To keep the average of $k_{\rm in}(t)$ over the period, 
the average of $(k^{\rm (L)}(t)+k^{\rm (R)}(t))f(t)$ must be kept zero.
One of the simplest choice is:
\be
 f(t) 
 = \frac{k_0}{2}\frac{x\cos\omega t
 +y\sin\omega t}{k^{\rm (L)}(t)+k^{\rm (R)}(t)}.
\ee

We show the protocols and the corresponding current in Fig.~3 of the main
body of the paper.
We set 
$(x,y)=(1.0,0.0)$ for the protocol 1,
$(x,y)=(0.0,1.0)$ for 2, 
$(x,y)=(0.0,-1.0)$ for 3, and 
$(x,y)=(-1.0,0.0)$ for 4.
The dynamical current is zero in all the protocols.

\section{Assisted adiabatic pumping}

\subsection{Choice of transition rates}

We obtained in the main body of the paper that 
the assisted adiabatic driving is achieved by using the replacement 
\be
 && k_{\rm in}(t)\to k_{\rm in}(t)-\dot{p}_{\rm out}(t), \\
 && k_{\rm out}(t)\to k_{\rm out}(t)+\dot{p}_{\rm out}(t),
\ee
where the dot denotes the time derivative.
This does not determine 
the decomposition of the left and right parts of the transition rates uniquely. 
We show in the following that 
we can find the ideal driving by 
\be
 && k_{\rm in}^{\rm (L)}\to k_{\rm in}^{\rm (L)}-\frac{k^{(\rm L)}_{\rm in}+k^{(\rm L)}_{\rm out}}{k}\dot{p}_{\rm out}, \\
 && k_{\rm in}^{\rm (R)}\to k_{\rm in}^{\rm (R)}-\frac{k^{(\rm R)}_{\rm in}+k^{(\rm R)}_{\rm out}}{k}\dot{p}_{\rm out}, \\
 && k_{\rm out}^{\rm (L)}\to k_{\rm out}^{\rm (L)}+\frac{k^{(\rm L)}_{\rm in}+k^{(\rm L)}_{\rm out}}{k}\dot{p}_{\rm out}, \\
 && k_{\rm out}^{\rm (R)}\to k_{\rm out}^{\rm (R)}+\frac{k^{(\rm R)}_{\rm in}+k^{(\rm R)}_{\rm out}}{k}\dot{p}_{\rm out},
\ee
where $k=k_{\rm in}^{\rm (L)}+k_{\rm in}^{\rm (R)}+k_{\rm out}^{\rm (L)}+k_{\rm out}^{\rm (R)}$.

The local dynamical current in Eq.~(\ref{Jdt}) is invariant under 
the above transformation.
The local geometric current is given by 
\be
 J_{\rm g}(t) =
 \frac{k^{\rm (R)}(t)}{k(t)}
 \frac{\diff}{\diff t}
 \left(p_{\rm out}(t)+\delta(t)\right).
\ee
$k^{\rm (R)}(t)/k(t)$ is invariant under the transformation.
$p_{\rm out}(t)$ is changed as 
\be
 p_{\rm out}(t)\to p_{\rm out}(t)+\frac{\dot{p}_{\rm out}(t)}{k(t)}.
\ee
We also see from the integral form in Eq.~(\ref{delta}) 
that $\delta(t)$ is changed as 
\be
 \delta(t) 
 &\to& 
 \delta(t)-\int_0^t \diff t'\,\frac{\diff}{\diff t'}\left(
 \frac{\dot{p}_{\rm out}(t')}{k(t')}\right)
 \e^{-\int_{t'}^{t} \diff t''\,(k_{\rm in}(t'')+k_{\rm out}(t''))} \no\\
 &=& 
-\frac{\dot{p}_{\rm out}(t)}{k(t)}
 +\frac{\dot{p}_{\rm out}(0)}{k(0)}
 \e^{-\int_{0}^{t} \diff t'\,(k_{\rm in}(t')+k_{\rm out}(t'))},
\ee
where we use the partial integration.
The last term is a decaying function and does not contribute to 
the current.
Then, we find 
\be
 J_{\rm g} \to
 \frac{k^{\rm (R)}(t)}{k(t)}\frac{\diff}{\diff t}p_{\rm out}(t),
\ee
which shows that the geometric current in the assisted system
including nonadiabatic effects 
is equal to the adiabatic current in the original system.

\subsection{Scaling}

Suppose that we have a time-independent $k_{\rm out}$ 
and want to keep that value after introducing the counterdiabatic term.
Using the time scaling
\be
 \tilde{t}(t)=\int_0^t\diff t'\,\left(1+\frac{\dot{p}_{\rm out}(t')}{k_{\rm out}}\right),
\ee
we obtain the master equation 
\be
 \frac{\diff}{\diff \tilde{t}}|\tilde{p}(\tilde{t})\rangle
 =\tilde{W}(\tilde{t})|\tilde{p}(\tilde{t})\rangle,
\ee
where 
\be
 \tilde{W}(\tilde{t})=\bmat{cc}
 -\tilde{k}_{\rm in}(\tilde{t}) & k_{\rm out} \\
 \tilde{k}_{\rm in}(\tilde{t}) & -k_{\rm out} \emat, 
\ee
and 
\be
 \tilde{k}_{\rm in}(\tilde{t})=\frac{1-\frac{\dot{p}_{\rm out}(t)}{k_{\rm in}(t)}}{1+\frac{\dot{p}_{\rm out}(t)}{k_{\rm out}}}k_{\rm in}(t).
\ee
Since $\tilde{t}$ is different from $t$,
the state at the scaled time $\tilde{t}$, $|\tilde{p}(\tilde{t})\rangle$,
is the adiabatic state at the original scale $t$.
We note that there is one-to-one correspondence between $t$ and $\tilde{t}$.
To keep the dynamical current invariant, 
we can use the decomposition 
$\tilde{k}_{\rm in}(t)=\tilde{k}_{\rm in}^{\rm (L)}(t)+\tilde{k}_{\rm in}^{\rm (R)}(t)$
where 
\be
 && \tilde{k}_{\rm in}^{\rm (L)} =
 \frac{k_{\rm in}-\dot{p}_{\rm out}}{k_{\rm out}+\dot{p}_{\rm out}}k_{\rm out}^{\rm (L)}
 +\frac{k_{\rm in}^{\rm (L)}k_{\rm out}^{\rm (R)}-k_{\rm out}^{\rm (L)}k_{\rm in}^{\rm (R)}}{\left(k_{\rm out}+\dot{p}_{\rm out}\right)^2}k_{\rm out}, \\
 && \tilde{k}_{\rm in}^{\rm (R)} = 
 \frac{k_{\rm in}-\dot{p}_{\rm out}}{k_{\rm out}+\dot{p}_{\rm out}}k_{\rm out}^{\rm (R)}
 -\frac{k_{\rm in}^{\rm (L)}k_{\rm out}^{\rm (R)}-k_{\rm out}^{\rm (L)}k_{\rm in}^{\rm (R)}}{\left(k_{\rm out}+\dot{p}_{\rm out}\right)^2}k_{\rm out}.
\ee

The obtained protocol is shown in Fig.~\ref{fig:s1} for a slow driving
and \ref{fig:s2} for a fast driving. 
The obtained current is shown in Fig.~4 in the main body of the paper.

\begin{widetext}
\begin{center}
\begin{figure}[th]
\begin{center}
\begin{minipage}[h]{0.45\textwidth}
\begin{center}
\includegraphics[width=1.\columnwidth]{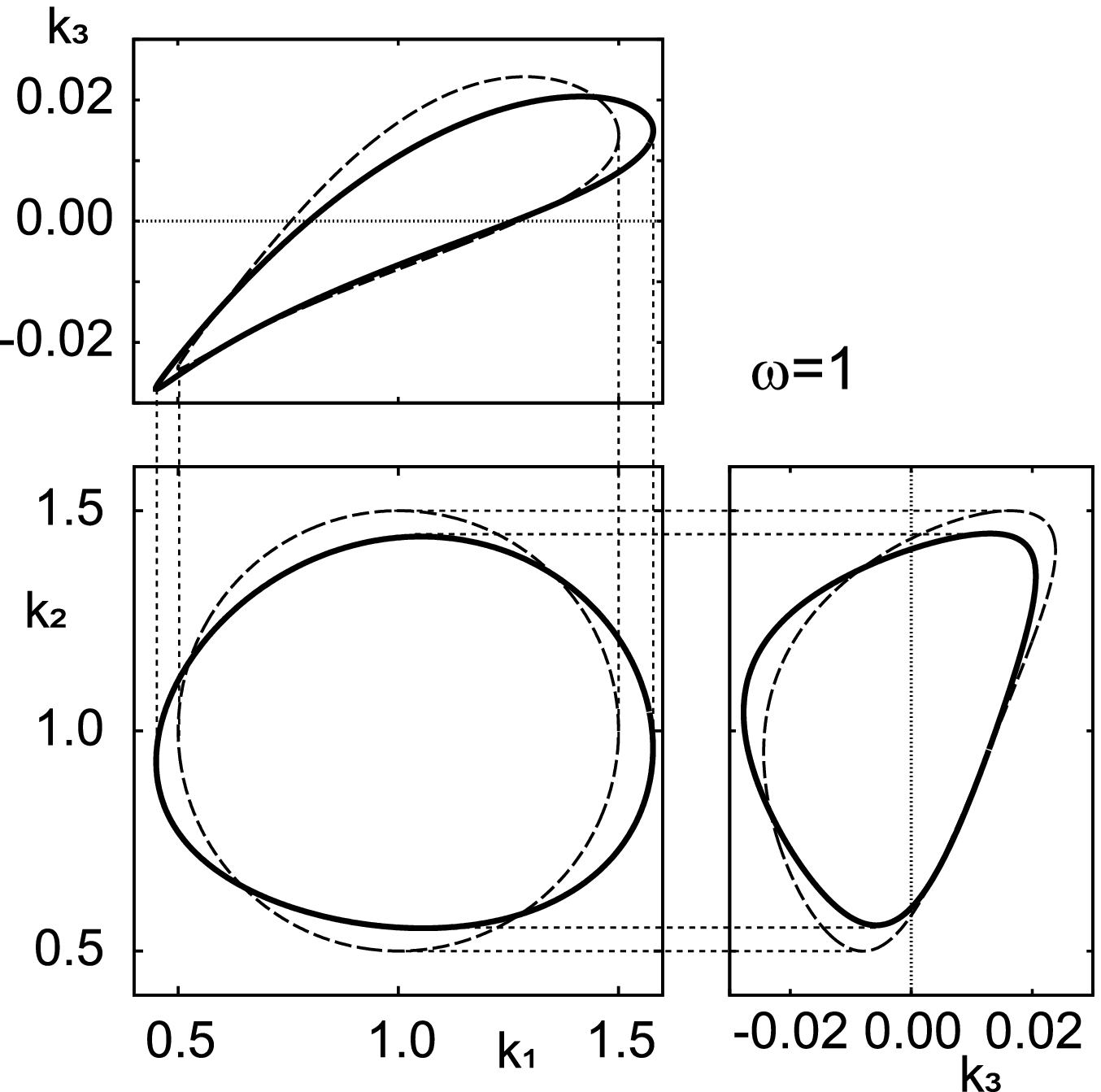}
\end{center}\end{minipage}\hspace*{0.1\textwidth}
\begin{minipage}[h]{0.45\textwidth}
\begin{center}
\includegraphics[width=1.\columnwidth]{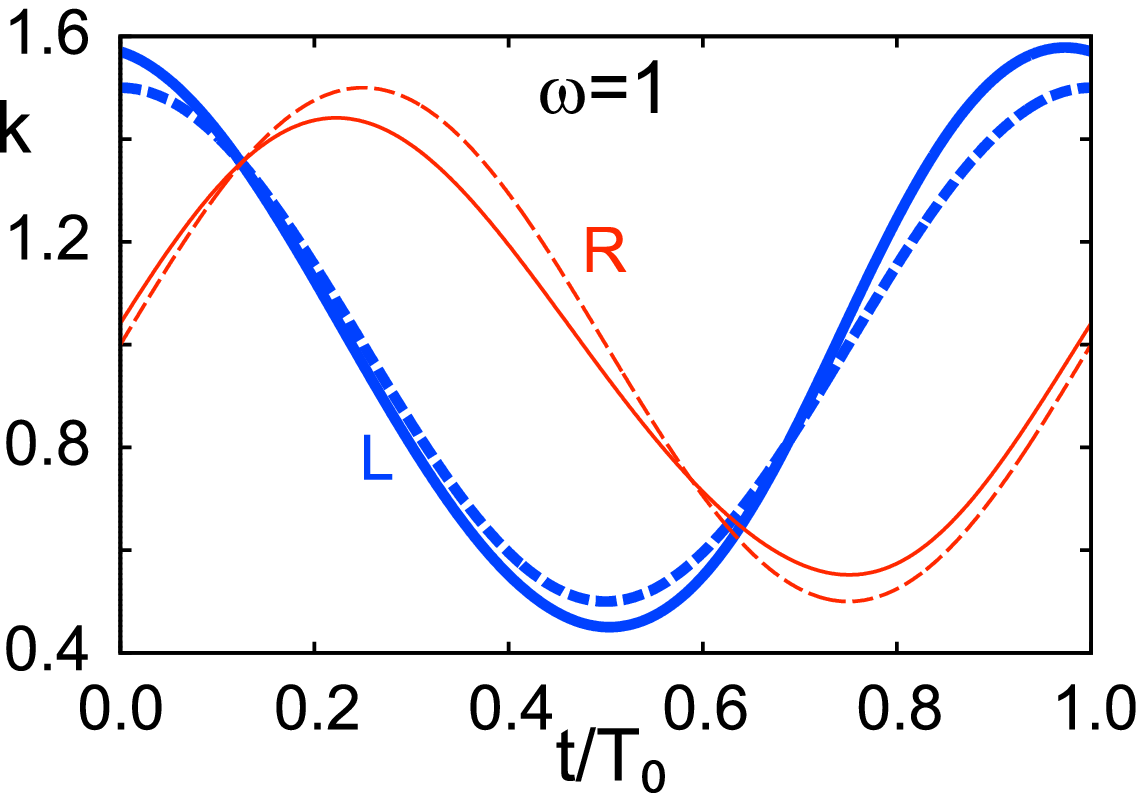}
\end{center}
\end{minipage}
\end{center}
\caption{
Protocols before/after the assist for $\omega=1.0$ . 
Top: Trajectories in parameter space.
Dashed lines represent trajectories of the original protocol
and solid lines of protocol with assist.
Bottom: Time dependence of the protocols.
Bold blue lines represent the left amplitude $k_{\rm in}^{\rm (L)}$
and thin red lines the right amplitude $k_{\rm in}^{\rm (R)}$.
Dashed lines represent protocols before assist and 
solid lines with assist.
}
\label{fig:s1}
\end{figure}
\end{center}
\begin{center}
\begin{figure}[ht]
\begin{center}
\begin{minipage}[h]{0.45\textwidth}
\begin{center}
\includegraphics[width=1.\columnwidth]{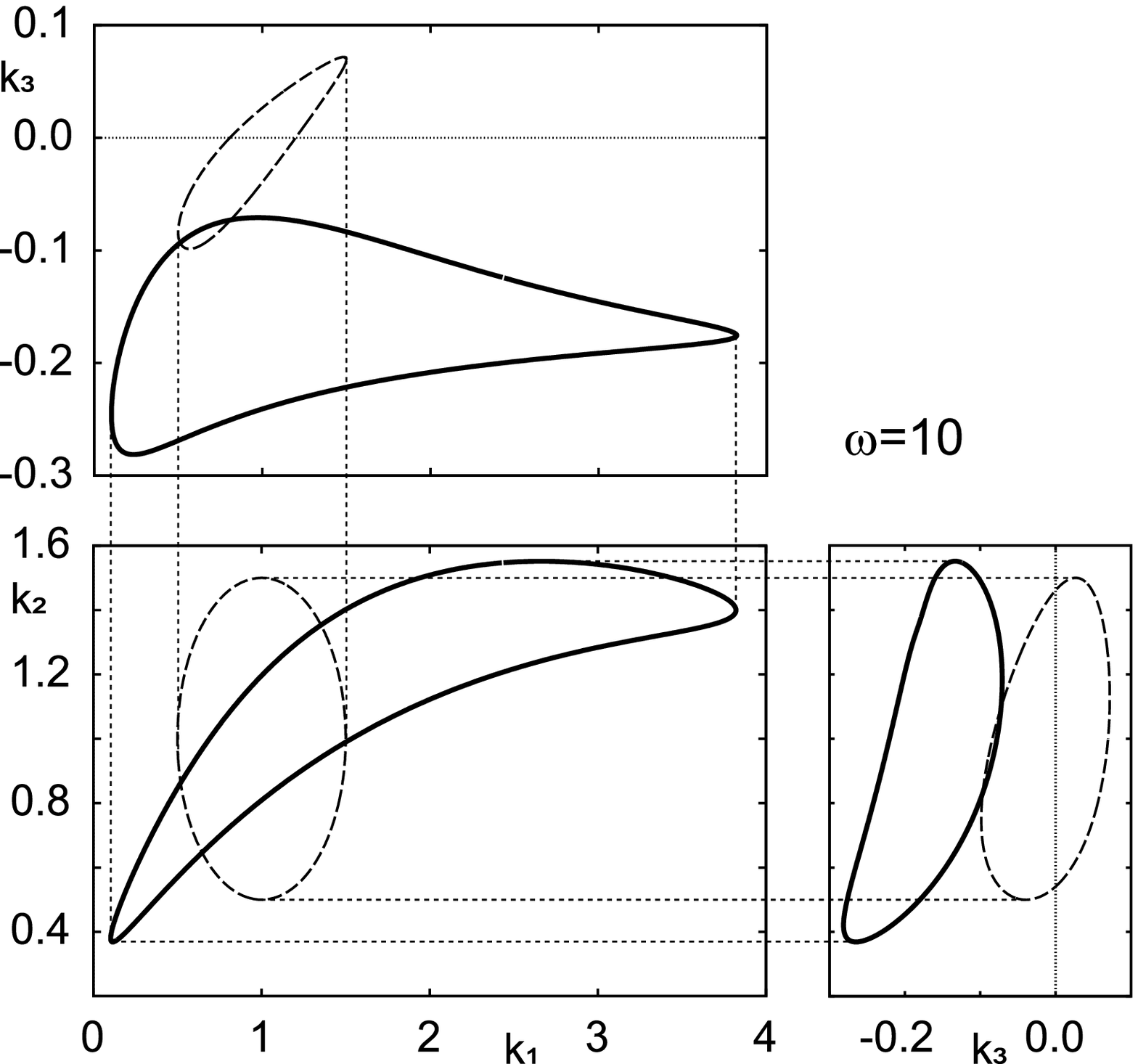}
\end{center}\end{minipage}\hspace*{0.1\textwidth}
\begin{minipage}[h]{0.45\textwidth}
\begin{center}
\includegraphics[width=1.\columnwidth]{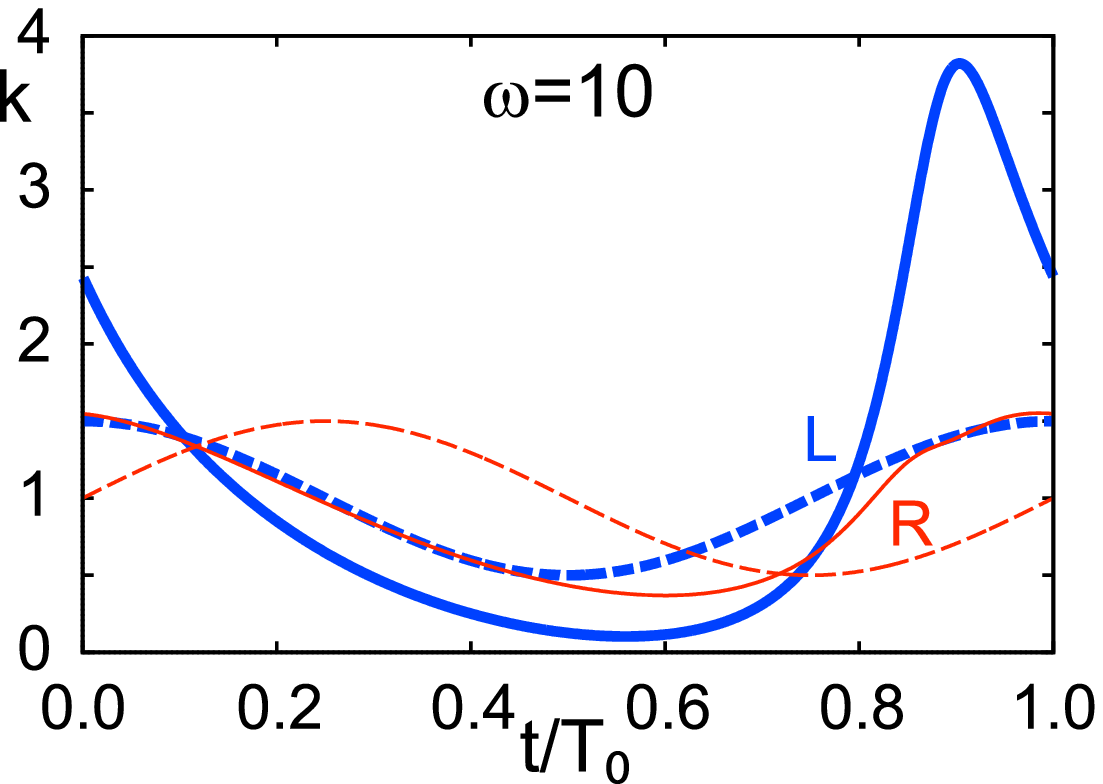}
\end{center}
\end{minipage}
\end{center}
\caption{
Protocols before/after the assist for $\omega=10.0$. 
}
\label{fig:s2}
\end{figure}
\end{center}
\end{widetext}

\subsection{Stability}

\begin{center}
\begin{figure}[ht]
\begin{center}
\includegraphics[width=1.\columnwidth]{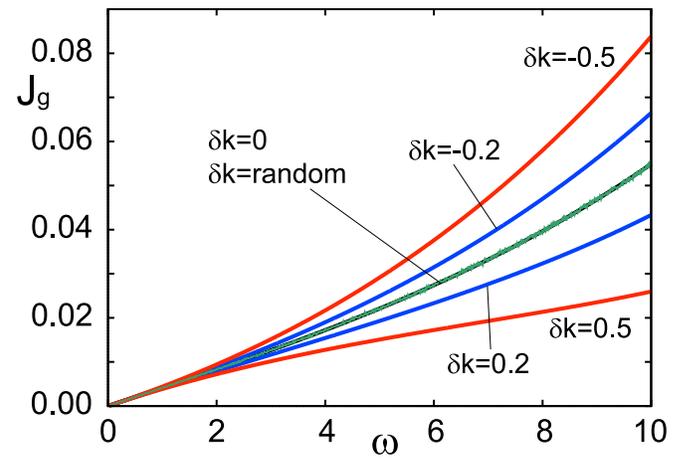}
\end{center}
\caption{
Stability of the driving.
For the plot ``$\delta k=\mbox{random}$'', we choose $\delta k(t)$ randomly
at each time between $-1$ and $1$.
It almost overlaps with the result at $\delta k=0$.
}
\label{fig:stab}
\end{figure}
\end{center}

To examine the stability of the assisted driving, we consider
small deviations from the ideal driving.
We put 
\be
 \tilde{k}_{\rm in}^{\rm (L)}\to \tilde{k}_{\rm in}^{\rm (L)}+\delta k, \\
 \tilde{k}_{\rm in}^{\rm (R)}\to  \tilde{k}_{\rm in}^{\rm (R)}-\delta k,
\ee
and choose $\delta k$ in several ways.
We plot the result of the geometric part of the current in Fig.~\ref{fig:stab}.
We see that the deviation becomes large at large frequencies 
compared to that at small frequencies.
The deviation is systematic and we do not find any instability.
When we choose $\delta k(t)$ randomly, 
the current is almost unchanged and the effect of randomness only 
gives very small oscillations.

\end{document}